\documentclass[namedreferences]{solarphysics}
\usepackage[optionalrh]{spr-sola-addons} 
\usepackage{graphicx}        
\usepackage{color}           
\usepackage{rotating}        
\usepackage{lscape}          
\usepackage{longtable}       
\usepackage{url}             

\newcommand{\aap}{    {\it Astron. Astrophys.}}

\newcommand{\apj}{    {\it Astrophys. J.}}
\newcommand{\apjl}{   {\it Astrophys. J. Lett.}}

\newcommand{\jgr}{    {\it J. Geophys. Res.}}

\newcommand{\solphys}{{\it Solar Phys.}}
 
\newcommand{\ssr}{    {\it Space Sci. Rev.}} 
\newcommand{\txw}{\textwidth}

\begin{document}

\begin{article}

\begin{opening}

\title{Near-Sun Flux Rope Structure of CMEs}

\author{H.~\surname{Xie}$^{1}$\sep
        N.~\surname{Gopalswamy}$^{2}$\sep
        O. C.~\surname{St.Cyr}$^{2}$      
       }
\runningauthor{Xie {\it et al.}}
\runningtitle{Flux Rope Structure of CMEs}

   \institute{$^{1}$ Department of Physics, The Catholic University of America, Washington, DC 20064, USA
                     email: \url{hong.xie@nasa.gov} \\ 
              $^{2}$ NASA Goddard Space Flight Center, Code 671, Greenbelt, MD 20771, USA
                     email: \url{nat.gopalswamy@nasa.gov} email: \url{chris.stcyr@nasa.gov} \\
             }

\begin{abstract}
We have used the Krall's flux-rope model 
(\citeauthor{Krall06}, \apj{} \citeyear{Krall06}, \textbf{657}, 1740) (KFR)
to fit  23 magnetic cloud (MC)-CMEs  and 30 non-cloud ejecta (EJ)-CMEs in the Living With a Star (LWS) Coordinated Data Analysis Workshop (CDAW) 2011 list. 
The KFR-fit results shows that the CMEs associated with MCs (EJs) 
have been deflected closer to (away from) the solar disk center (DC), likely 
by both the intrinsic magnetic structures inside an active region (AR) and ambient magnetic 
structures  ({\it e.g.} nearby ARs, coronal holes, and streamers, {\it etc.}). 
The mean propagation latitudes and longitudes of the EJ-CMEs (18$^\circ$, 11$^\circ$) were larger than those of 
the MC-CMEs (11$^\circ$, 6$^\circ$) by 7$^\circ$ and 5$^\circ$, respectively. 
Furthermore, the KFR-fit widths showed that the MC-CMEs are wider than the EJ-CMEs.
The mean fitting face-on width and edge-on width of the MC-CMEs (EJ-CMEs) were 
87 (85)$^\circ$  and 70 (63)$^\circ$, respectively.
The deflection away from DC and narrower angular widths of the EJ-CMEs have caused the observing spacecraft to pass over only their flanks and miss the central flux-rope structures.
The results of this work support the idea that all CMEs have a flux-rope structure.
\end{abstract}
\keywords{Coronal Mass Ejections, Initiation and Propagation}
\end{opening}

\section{Introduction}

In recent years, a great deal of research in both modeling and observations 
(see, {\it e.g.}, \opencite{Chen97}, \opencite{Dere99}, \opencite{Gibson00}, \opencite{Krall01}, 
\opencite{Crema04}, \opencite{Krall06}, \opencite{Thern06}, \opencite{Krall07}) has been focused on 
coronal mass ejections (CMEs) having the ``three-part" morphology, namely, 
a bright front, a dark void and a bright core of prominence material \cite{Illing85}. 
Concave-outward trailing features were noted in the {\it Solar Maximum Mission} (SMM) coronagraph images (\opencite{Illing85}, \opencite{Burke93}), but the high-resolution {\it Large Angle and Spectrometric COronagraph} (LASCO) observations revealed that these three-part CMEs often create 
the appearance of a helical or flux-rope (FR) structure. 

The interplanetary (IP) counterpart of CMEs are called ICMEs. 
They are featured with high magnetic fields, low ion temperatures,
high  alpha/proton density ratios and occasionally bidirectional streaming of 
electrons and ions ({\it e.g.}, \opencite{Gosli90}).  
Depending on whether they have a smooth rotating magnetic field, 
{\it i.e.}, a signature of flux rope, 
and fulfill other conditions,
ICMEs are further classified into: a) magnetic clouds (MCs), and b) non-cloud ejecta (or simply ejecta)(EJs)
(see, {\it e.g.}, \opencite{Burla81}, \opencite{Gosli90}). 
\inlinecite{Gopal06} suggested that all CMEs have magnetic FR structures, 
but their propagation directions determine whether they can be seen {\it in-situ}. 

 

Based on the coronagraphic observations, simple FR models of CMEs using a torus geometry  have been 
developed by various authors.  \inlinecite{Krall06} (hereafter KS06) described a FR model as 
having an elliptical curved axis with a circular cross-section of varying radius along the axis and the width (minor diameter) being narrowest at the foot-points on the solar surface. 
It was shown that the KS06 FR model (KFR) geometry reproduced the statistical measures 
(average angular widths) of a subset of FR-like CMEs observed by LASCO \cite{Stcyr04}.  
\inlinecite{Thern06} used the graduated cylindrical shell (GCS) model, a FR model 
with a conical curved axis, and demonstrated that the GCS model fits the CME morphologies 
for 34 FR-like CMEs selected from \inlinecite{Crema04}. A more recent study \cite{Krall07} 
extended the work of KS06 by comparing the FR model synthetic images to 111 limb CMEs 
observed by SMM \cite{Burke04}. Their results suggested that 
the FR morphology can be applied not only to FR-like CMEs but also to the 
general population of CMEs. 

In this work, we apply the KFR to the 2011 Living With a Star (LWS) Coordinated Data Analysis
Workshop (CDAW) CME list to determine the radial speeds, angular widths, and the propagation 
directions  of the CMEs.
We studied 53 shock-driving CMEs during Solar Cycle 23 whose 
source regions are located within E15$^\circ$ and W15$^\circ$. We have excluded complex event number 3 with multiple solar sources: an eruptive prominence (EP) (N45W10), active region (AR) 8115 (N32W32), and AR 8113 (N25W40).
To zeroth order, it is supposed that MCs associated CMEs (MC-CMEs) come from the disk center, 
ejecta-CMEs come from intermediate longitudes, 
and driverless shocks come from near the limb ({\it e.g.}, \opencite{Gopal06}).
Since all the CDAW CMEs originate raletively from the disk center, it is expected 
that they are observed as MCs at Earth, assuming that all CMEs have FR structures and erupt radially.
However, for the 53 CDAW events, 23 events are associated with MCs and 30 events are EJs. 
One likely reason is the non-radial eruption of CMEs 
(see, {\it e.g.}, \opencite{Stcyr00}, \opencite{Gopal09}, \opencite{Xie09}).
In this paper, we are looking into the characteristics that distinguish two CME populations: 
MC-CMEs and EJ-CMEs.
By performing a detailed investigation of CME origins and propagations, 
we hope to answer the CDAW focusing question:  Do all CMEs have flux rope structure, 
but sometimes they are not observed so because of geometry (the observing spacecraft 
does not pass through the flux rope) or do some CMEs have inherently non-flux rope structure?

The rest of the paper is organized as follows. Section 2 describes data selection 
and detailed FR model fitting procedures. Section 3 presents the fitting results and
statistical analysis. Finally, the summary and conclusions are presented in Section 4. 

\section{Data and Model}
The CDAW 2011 list is a subset of the list in \inlinecite{Gopal10}. 
This subset was selected based on two criteria: 1) the CMEs originate from close to the central meridian 
(E15$^\circ \leq$ source longitude $\leq$ W15$^\circ$), 
2)  the CMEs are associated with shock-driving interplanetary CMEs (ICMEs). 
The CDAW 2011 list consists of 23 MC-CMEs and 30 EJ-CMEs.  

\begin{figure}
 \center\includegraphics[width= .9\txw,angle=0]{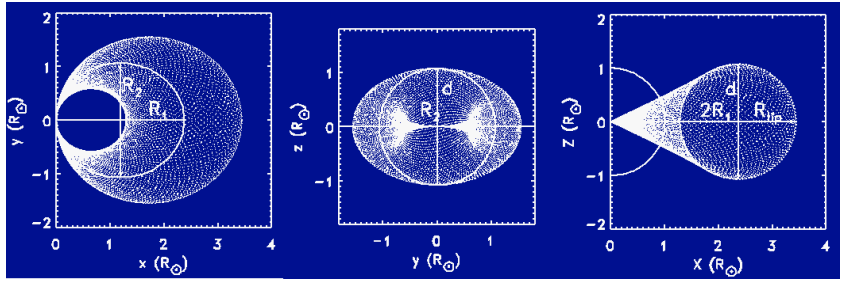}
 \caption{Illustrative plot of the flux-rope model morphology. From left to right: 
          broadside (face-on), top, and edge-on views of a flux-rope.}\label{F-FRmodel}
 \end{figure}

To determine the radial speeds, angular widths, and propagation directions of 
the CMEs, we applied the KS06 FR model fit (KFR-fit) to LASCO C2 and C3 images. 
The KS06 model is also called the elliptical FR model, which assumes that the FR has an elliptical axis with varying radial circular cross-section. 
Figure~\ref{F-FRmodel} gives the broadside (face-on), top, and edge-on views of the FR model,
with apex pointing to the west limb. 

The geometry of the flux-rope can be described by two parameters: the ratio 
of the semi-minor to semi-major axis of the ellipse, $\lambda_{\epsilon} = R_2/R_1$ 
and the axial aspect ratio, $\Lambda_{\alpha} = 2R_1/d$, where $R_1$, $R_2$ , 
and $d$ are semi-major axis, semi-minor axis and width of the flux rope at 
its apex, as shown in Figure 1. The orientation of the flux rope is defined 
by three angles: latitude $\lambda$, longitude $\phi$, and tilt angle $\alpha$, 
where the tilt angle is the rotation angle clockwise with respect to East direction.

We used an iterative method to parameterize the flux rope model.  
First, we chose initial test parameters of the model based on the 
coronagraphic observations;  we then iteratively adjusted the test parameters 
until the best fit of the FR model to LASCO images was obtained. 
The fitted CME radial speed is given by $V_{CME} = \Delta (R_{tip})/dt$, where $R_{tip}$ is 
the radial distance from the origin to the apex of the FR.
The widths of the CME are given by: 
$\omega_{edge} = 2 \times tg^{-1}(0.5/\Lambda_{\alpha}), 
\omega_{broad} = 2 \times tg^{-1}(\lambda_{\epsilon})$, 
where $\omega_{edge}$ and $\omega_{broad}$ are the widths of the CME from 
edge-on and face-on views, respectively.

\begin{figure}
 \center\includegraphics[width=.75\txw,angle=0]{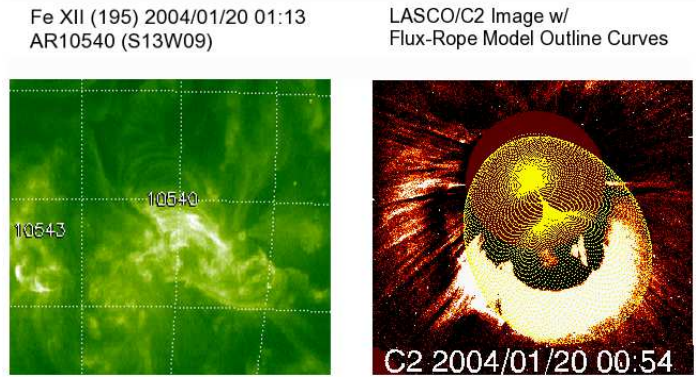}
 \caption{An example of the flux-rope model fit for the 2004 January 20 CME. 
          (Left) EIT 195 $\AA$ image showing the flare post-eruption arcade at 01:13 UT.
          (Right) LASCO/C2 image at 00:54 UT superimposed with the flur-rope model outline curves (yellow curves).
 }\label{F-FRfit}
 \end{figure}

Figure~\ref{F-FRfit} shows an example of the model fit for the 2004 January 20 CME. 
The CME was a halo CME associated with a C5.5 soft X-ray flare recorded by GOES 
in AR 10540 (S13W09), with a peak at 00:45 UT. 
Left panel of Figure ~\ref{F-FRfit} is the {\it Extreme Ultraviolet Imaging Telescope} (EIT) 195 $\AA$ image, 
which shows the post-eruption arcade titled $\sim$52$^\circ$ relative to the E-W direction at 01:13 UT. 
The right panel shows the model outline curves (yellow curves) superimposed on the LASCO/C2  
image at 00:54 UT.  The fitting gave a CME radial speed $V_{CME}$ =  1441 km/s,  
$\omega_{broad} = 90^\circ$ , and $\omega_{edge} = 71^\circ$; the best-fit for the propagation 
direction was ($\lambda,\phi,\alpha$) = (-25$^\circ$, 10$^\circ$, 60$^\circ$).  
The fitting results showed  that the CME erupted non-radially and was deflected from S13 to S25; 
the longitude and tilt angle of the KFR-fit direction were relatively consistent with 
the CME source location.  

\section{Statistical analysis and results} 

Table~\ref{T-FRfit} summarizes the flux-rope fitting results for the 53 CMEs. 
Columns 1 - 8 are the event number, shock date, time, ICME type,  CME date, first 
appearance time at LASCO C2, sky-plane speed, and  source location identified based on solar 
surface activities: flare,  EIT wave and dimming, eruptive prominence or disappearing filament.  
Columns 9 - 13 are outputs of the KFR-fit edge-on width and face-on (broadside) width, 
radial speed, propagation direction and tilt angle.

\subsection{Spatial relationship between CME source locations and propagation directions}

 \begin{figure}
    \mbox{
    \includegraphics[width=0.47\txw]{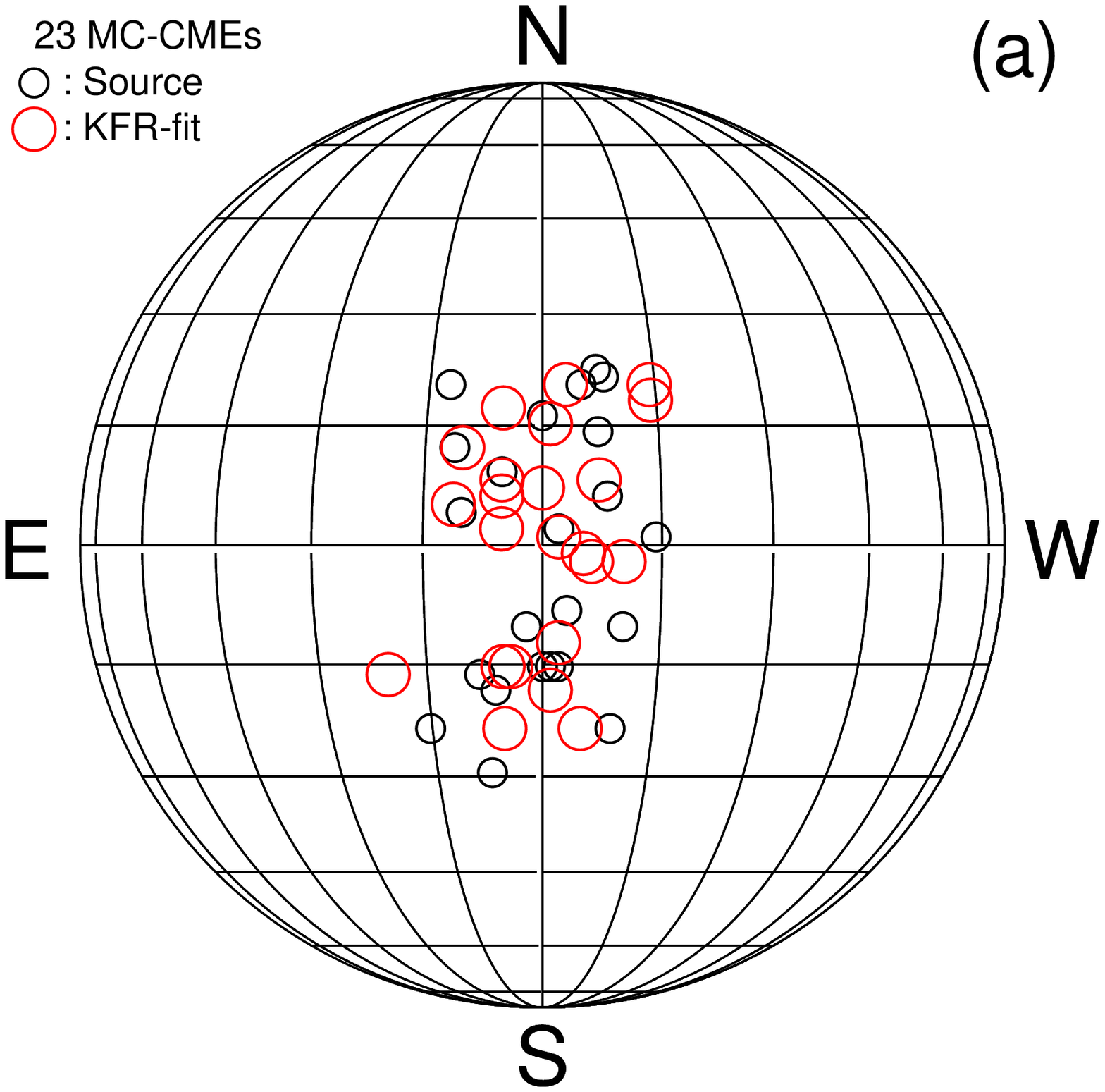}
    \includegraphics[width=0.47\txw]{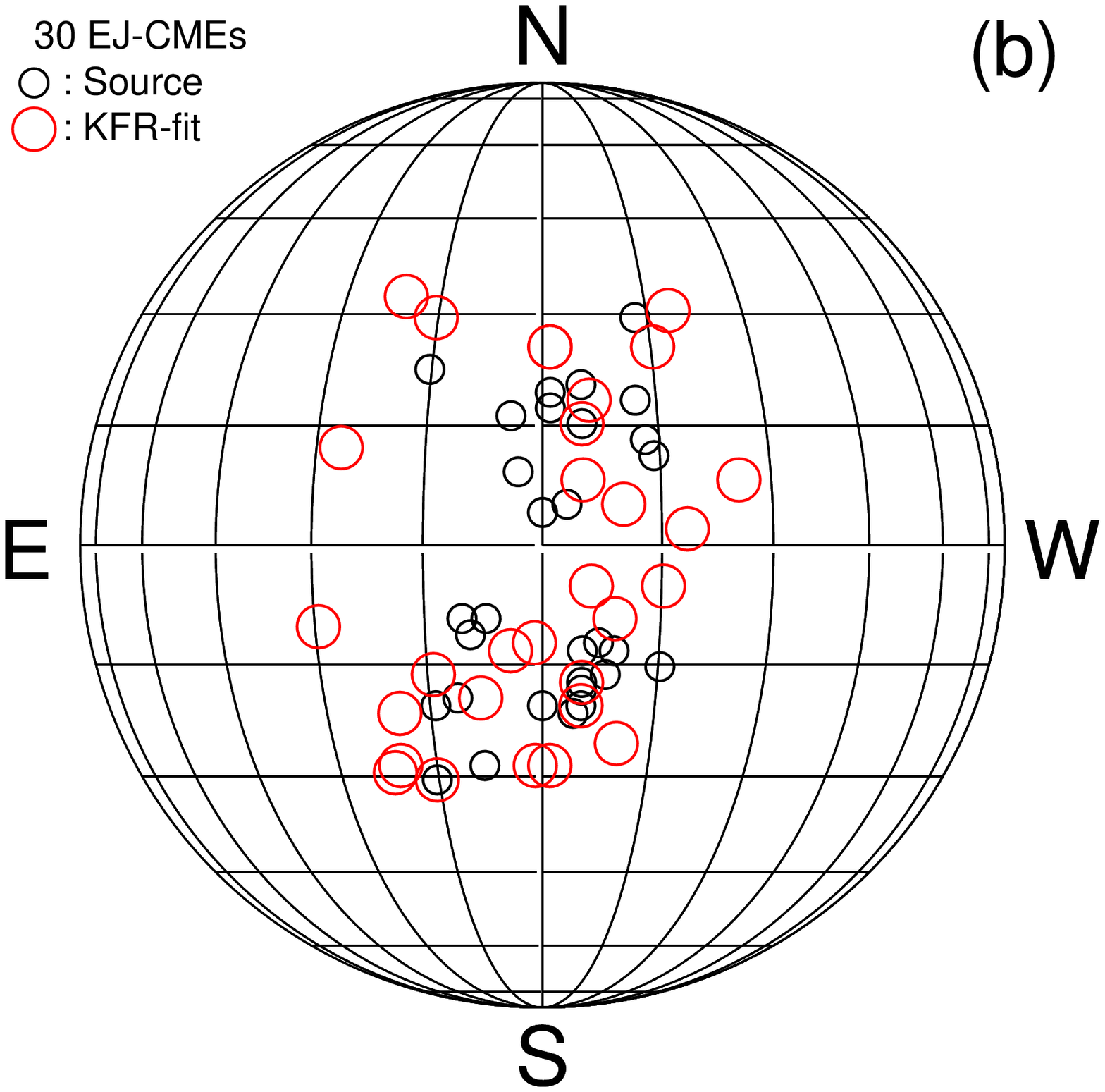}
   }\par
 \caption{Comparison between the source locations (black circles) and the KFR-fit propagation 
          directions (red circles) for a) 23 MC-CMEs and b) 30 EJ-CMEs.}
 \label{F-hgrid}   
 \end{figure}

\begin{figure} 
 \center\includegraphics[width= .9\txw,angle=0]{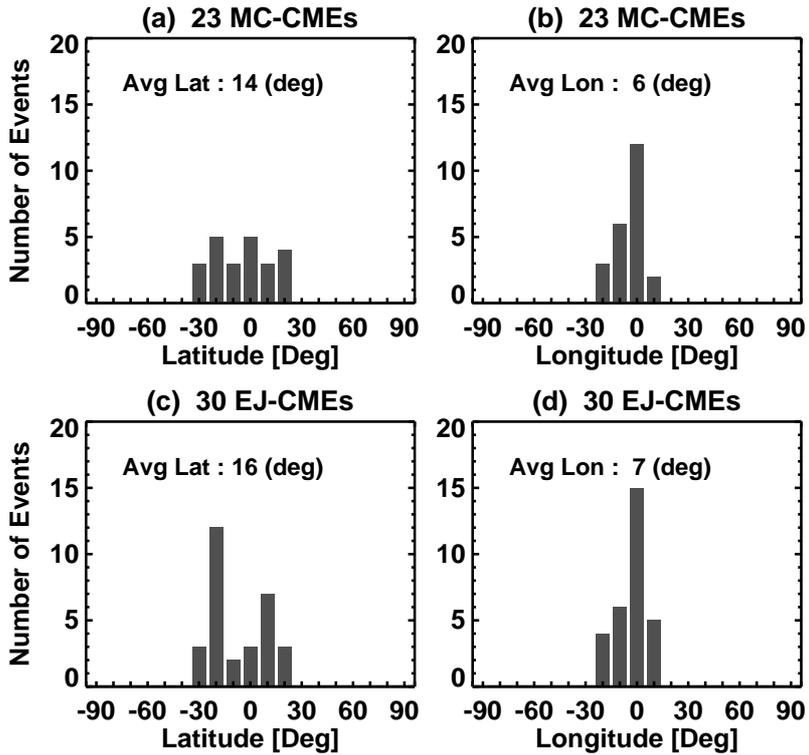}
 \caption{Histograms of the source locations: a) latitude and b) longitude for 23 MC-CMEs,
          c) latitude and d) longitude for 30 EJ-CMEs.}
 \label{F-scdir}   
 \end{figure}

\begin{figure} 
 \center\includegraphics[width= .9\txw,angle=0]{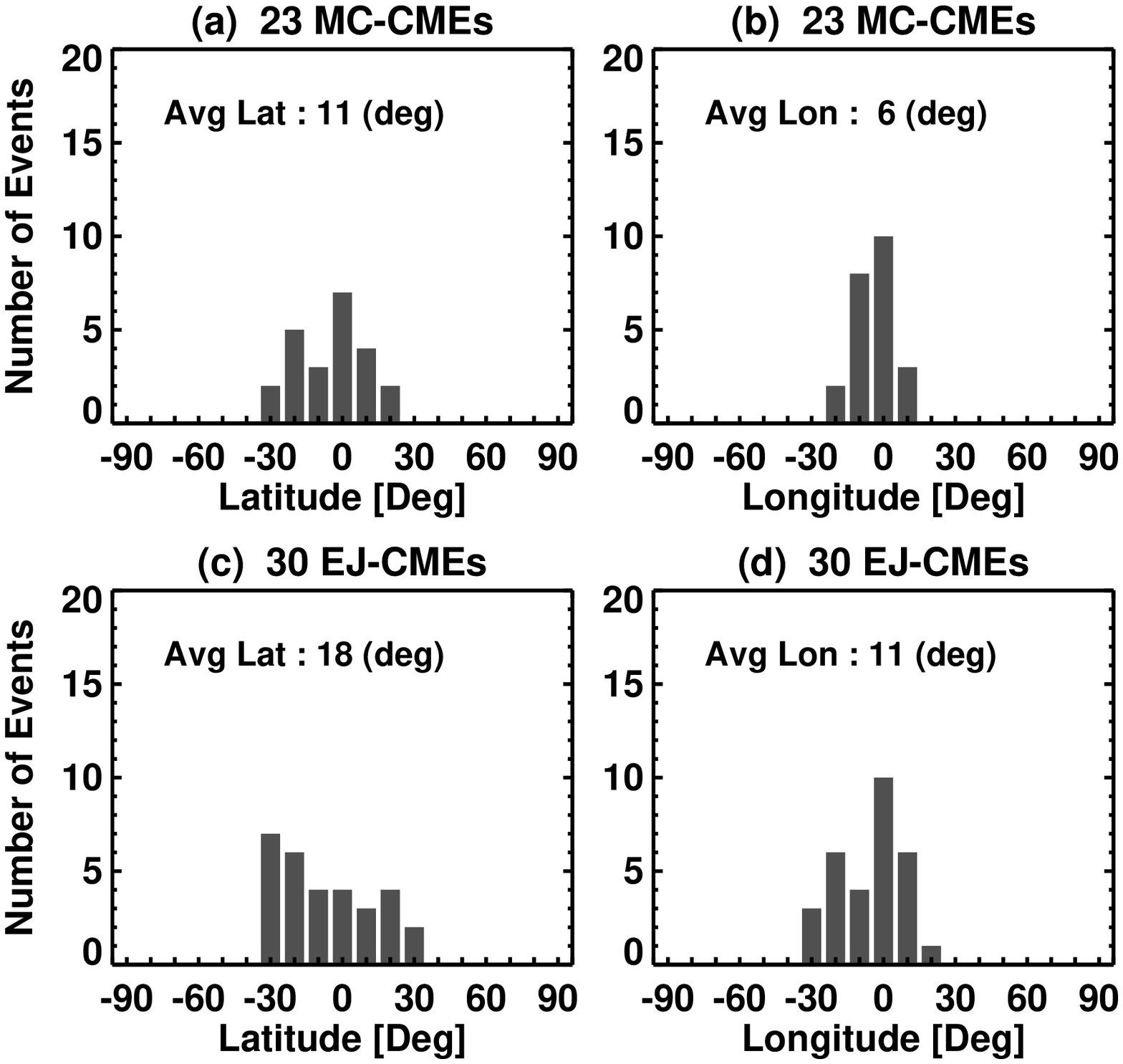}
 \caption{Histograms of the KFR-fit propagation directions: a) latitude and b) longitude for 23 MC-CMEs,
          c) latitude and d) longitude for 30 EJ-CMEs.}
 \label{F-frdir}   
 \end{figure}

Previous studies have shown that CMEs may erupt non-radially and be deflected by the ambient magnetic 
environment such as coronal holes ({\it e.g.} \opencite{Gopal09}, \opencite{Amaal12}), streamers,   
and nearby ARs or complex intrinsic structures of associated ARs (e.g., \opencite{Xie09}). 
Figure~\ref{F-hgrid} shows the CME source locations (black circles) and the KFR-fit propagation directions (red circles) 
for a) 23 MC-CMEs and b) 30 EJ-CMEs. 
From Figure~\ref{F-hgrid}a we can see that both the CME source locations and the KFR-fit propagation directions for the MC-CMEs are relatively clustered  close to DC.
Histograms of the source locations (Figures~\ref{F-scdir}a and b) show that the mean latitude and longitude of the source locations for the MC-CMEs are (14$^\circ$°,6$^\circ$°); 
and the mean latitude and longitude of the KFR-fit propagation directions are (11$^\circ$°,6$^\circ$°) (Figure~\ref{F-frdir}a and b).
While from Figure~\ref{F-hgrid}b, we see that the KFR-fit propagation directions 
of the EJ-CMEs are relatively scattered away from DC, compared to their source locations, indicating some degree of CME deflection. 
These deflections of the EJ-CMEs are also shown in Figure~\ref{F-frdir}.
In Figure~\ref{F-scdir} the mean latitude and longitude of the source locations for the EJ-CMEs are (16$^\circ$,7$^\circ$),
but the mean latitude and longitude of the KFR-fit propagation directions for the EJ-CMEs are (18$^\circ$,11$^\circ$) (Figure~\ref{F-frdir}c and d). 
 
\begin{figure} 
 \center\includegraphics[width= .9\txw,angle=0]{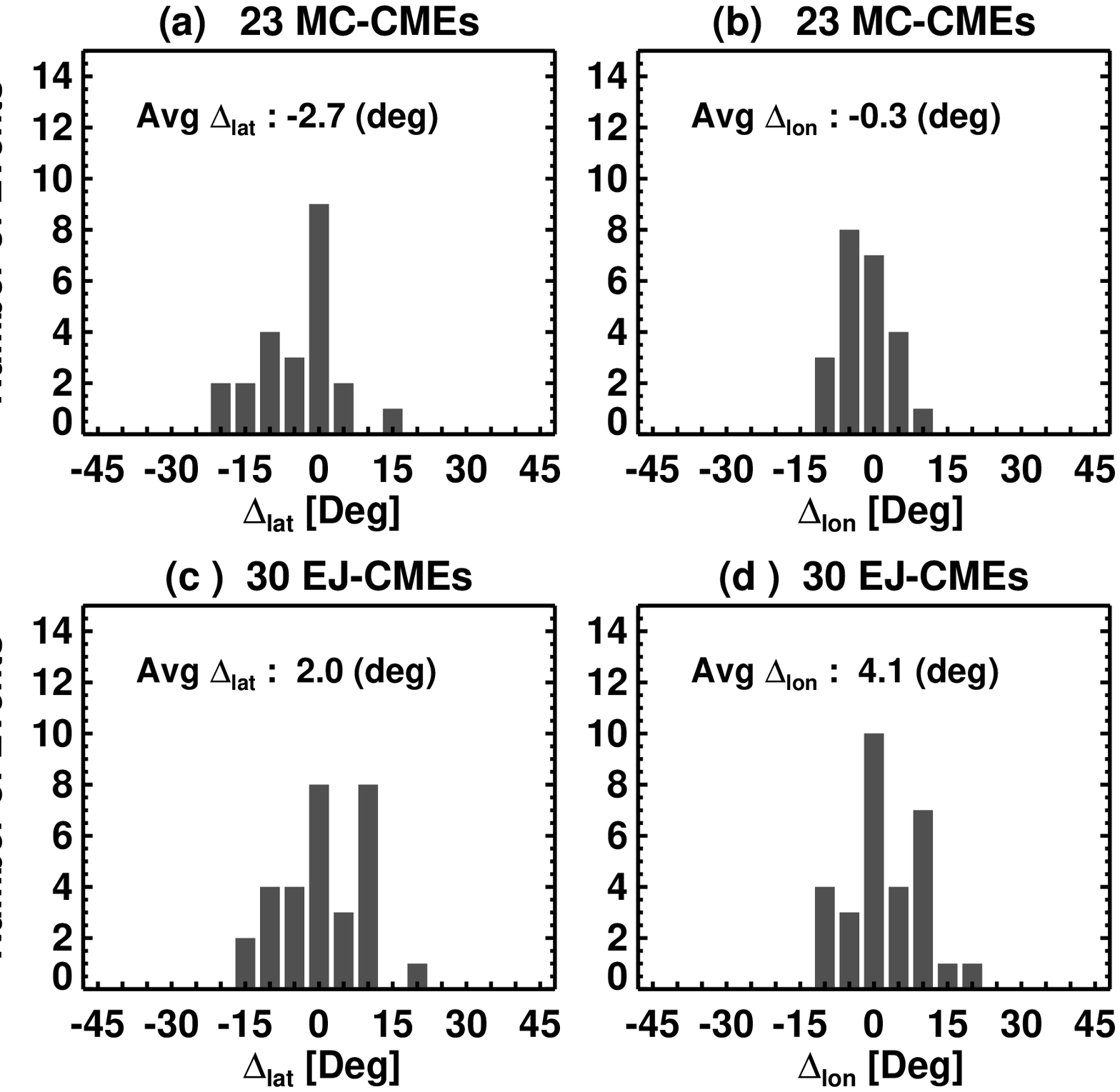}

 \caption{Histograms of latitudinal difference between the CME source locations 
           and propagation directions $\Delta_{lat}$ for a) 23 MC-CMEs and c) 30 EJ-CMEs,
           and histograms of longitudinal difference $\Delta_{lon}$ for b) 23 MC-CMEs and d) 30 EJ-CMEs.
          }\label{F-histdelt}
 \end{figure}

Figure~\ref{F-histdelt} plots histograms of: a) $\Delta_{lat}$ and b) $\Delta_{lon}$ for 23
MC-CMEs, and c) $\Delta_{lat}$ and d) $\Delta_{lon}$  for 30 EJ-CMEs,
where $\Delta_{lat}$ = $|Lat_{fr}|$ - $|Lat_{sc}|$ and  $\Delta_{lon}$ = $|Lon_{fr}|$ - $|Lon_{sc}|$ are the latitudinal and longitudinal differences
between CME source locations ($Lat_{sc}$, $Lon_{sc}$) and KFR-fit propagation directions ($Lat_{fr}$, $Lon_{fr}$). 
The mean  $\Delta_{lat}$ and $\Delta_{lon}$ for the MC-CMEs are -2.7$^\circ$ and -0.3$^\circ$,
and the mean  $\Delta_{lat}$ and $\Delta_{lon}$ for the EJ-CMEs are 2.0$^\circ$ and 4.1$^\circ$.
Since positive (negative) values of $\Delta_{lat}$ and  $\Delta_{lon}$ indicate that the CMEs were deflected away from (towards) DC, both Figure~\ref{F-frdir} and Figure~\ref{F-histdelt} 
suggest that the EJ-CMEs were deflected farther from DC while the MC-CMEs  were deflected closer to DC during their eruption and propagation near the Sun.


\subsection{Comparison of angular widths between MC-CMEs and EJ-CMEs} 

\begin{figure} 
 \center\includegraphics[width= .9\txw,angle=0]{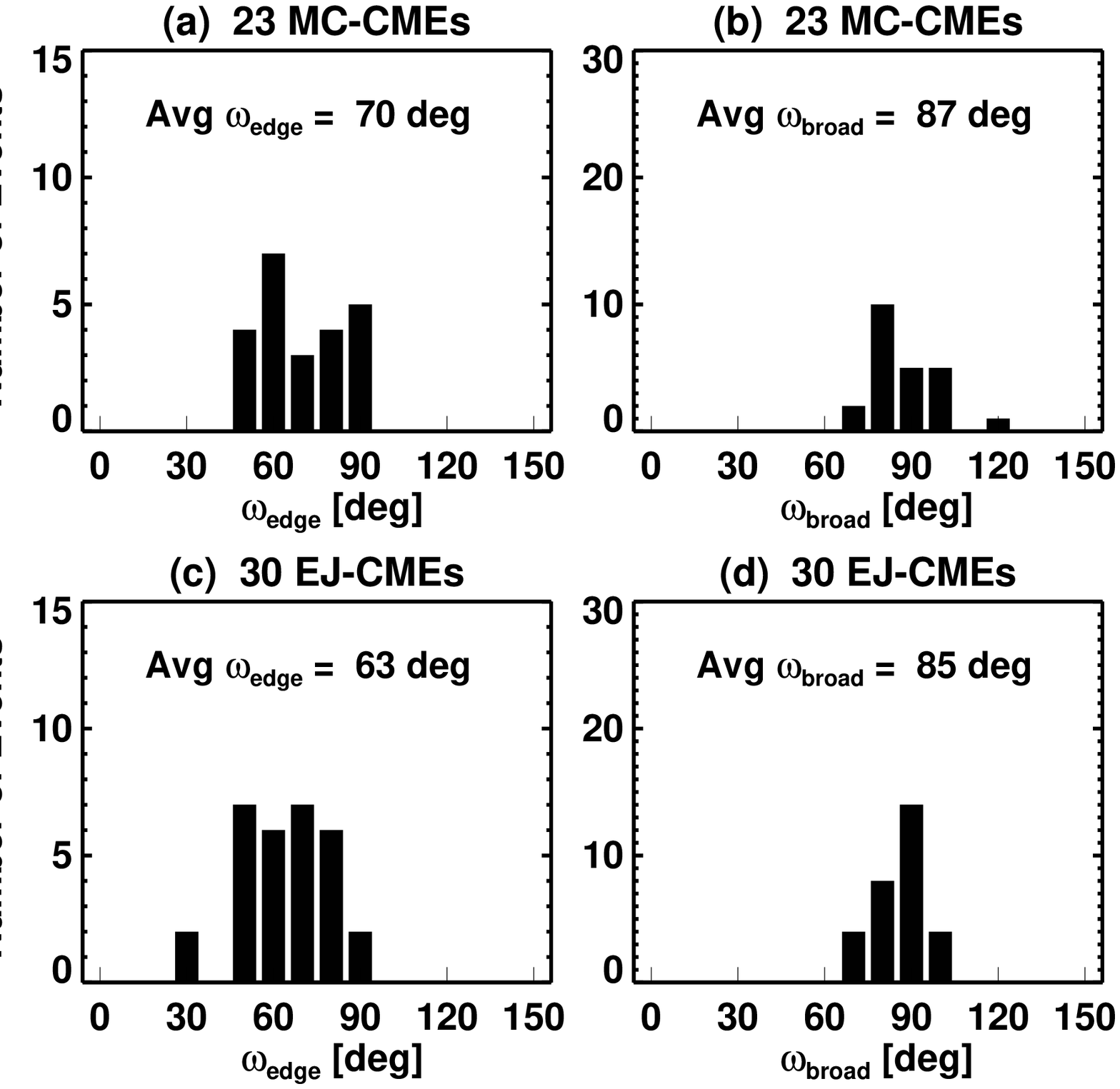}
 \caption{Histograms of KFR-fit widths: a) edge-on width $\omega_{edge}$ and b) broadside  width $\omega_{broad}$ for 23 MC-CMEs,
 c) $\omega_{edge}$ and  d) $\omega_{broad}$ for 30 EJ-CMEs.}
 \label{F-histw}
 \end{figure}

Figure~\ref{F-histw} presents histograms of a) $\omega_{edge}$ and b) $\omega_{broad}$ 
for 23 MC-CMEs, and histograms of c) $\omega_{edge}$ and d) $\omega_{broad}$ for 30 EJ-CMEs,
where $\omega_{edge}$ and $\omega_{broad}$ are the KFR-fit widths of the CMEs from 
edge-on and face-on views, respectively. 
The mean broadside (face-on) widths are 87$^\circ$ and 85$^\circ$ for the MC-CMEs and EJ-CMEs, 
and the mean edge-on widths are 70$^\circ$ and 63$^\circ$ for the MC-CMEs and EJ-CMEs, respectively.
The MC-CMEs are wider than the EJ-CMEs by  2 $^\circ$  in the mean broadside width and 7 $^\circ$ in the mean edge-on width. This indicates that the MC-CMEs are not only deflected toward DC, but also are larger in widths, thus 
their central FR structures are more likely observed in-situ.

\subsection{Comparison of radial speeds between MC-CMEs and EJ-CMEs } 

\begin{figure} 
 \center\includegraphics[width= .9\txw,angle=0]{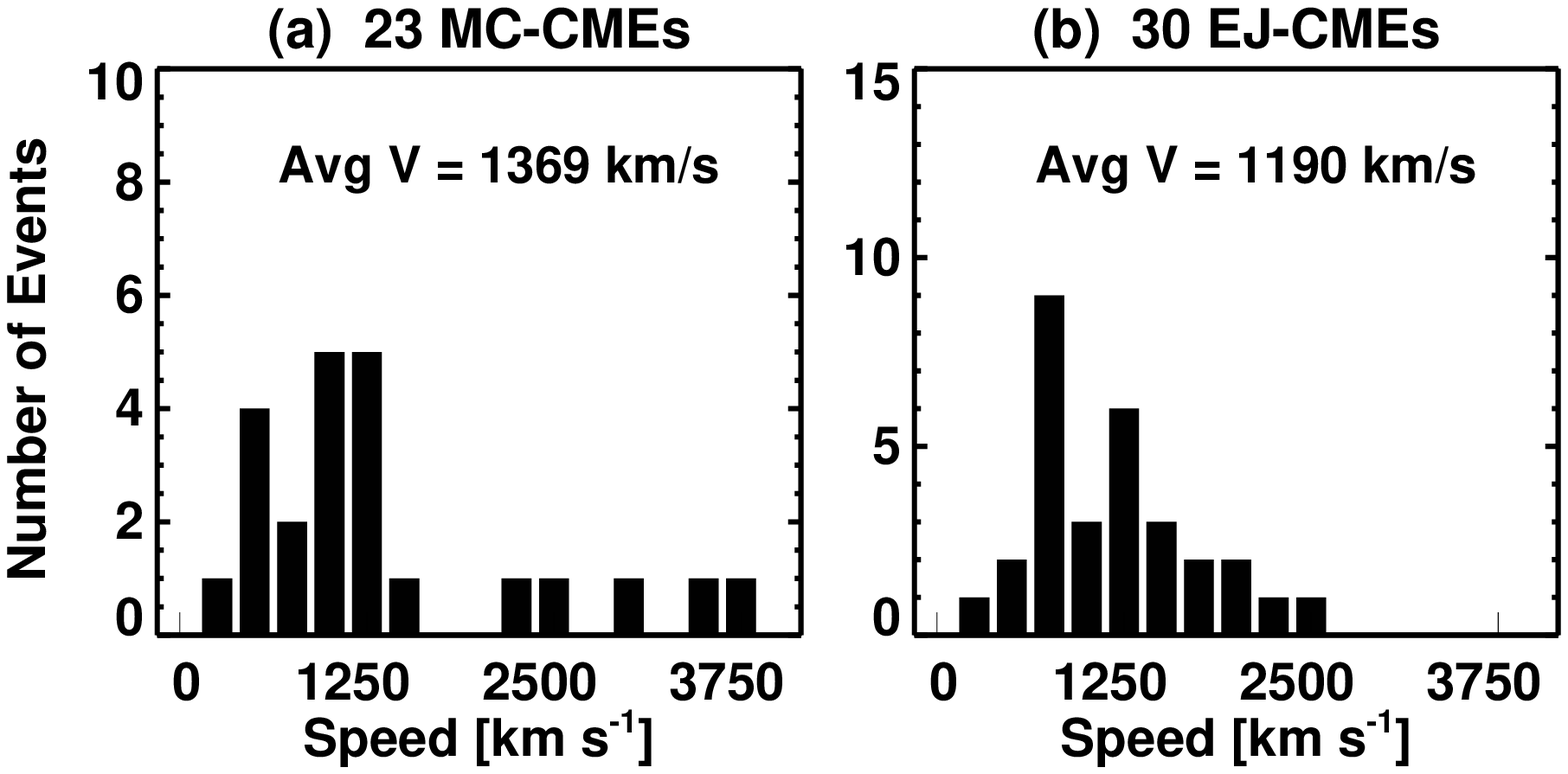}
 \caption{Histograms of KFR-fit radial speed $V_{cme}$ for a) 23 MC-CMEs and b) 30 EJ-CMEs.}
 \label{F-histvcme}
 \end{figure}

Figure~\ref{F-histvcme} shows histograms of radial speeds for: a) 23 MC-CMEs and b) 30 EJ-CMEs. 
The mean radial speeds are 1369 $km s^{-1}$ and 1190  $km s^{-1}$ for the MC-CMEs and EJ-CMEs, respectively.
The MC-CMEs are faster than EJ-CMEs by 121 $km s^{-1}$ concerning their mean radial speeds.
This might be one of other factors affecting the observed ICME properties. 
It is expected that the FR structures in the slow CMEs become more distorted by their interactions 
with the background solar wind 
and/or with other CMEs ({\it c.f.}, \opencite{Kim12}) during their propagation.


\section{Summary and Conclusion} 

We studied a set of the CDAW 2011 CMEs which consisted of 23 MC-CMEs and 30 EJ-CMEs.
These two groups of CMEs originated from similar source locations, with mean latitude and longitude of the MC-CMEs (EJ-CMEs) of 14 (16)$^\circ$ and 6 (7)$^\circ$, respectively.
We applied the KFR-fit to determine
the CME radial speeds, angular widths, and propagation directions.
The KFR-fit results have revealed that the properties of these two groups of CMEs 
showed no characteristic differences.

However, there exist distinguishing features between the two groups
in terms of their propagation directions and angular widths.
It is found that the EJ-CMEs tend to propagate in higher latitudinal and longitudinal directions. 
The mean propagation latitude and longitude of the EJ-CMEs were larger than those of 
the MC-CMEs by 7$^\circ$ and 5$^\circ$ (Figure~\ref{F-frdir}), respectively, 
The likely reasons of the CME non-radial eruption are the complex intrinsic structures of the associated ARs, deflections from the ambient magnetic structures such as coronal holes, streamers, and nearby ARs. 
It is shown in Figure~\ref{F-histdelt} that the EJ-CMEs were deflected away from DC, 
while the MC-CMEs were deflected towards DC. 
Similar results are also found in \inlinecite{Gopal09} and \inlinecite{Amaal12}, where they studied the interaction  
of CMEs with coronal holes. \inlinecite{Gopal09} showed that some fast and wide CMEs from DC were deflected away from the Sun-Earth line 
and the spacecraft at L1 only observed driveless-like shocks.  
\inlinecite{Amaal12} calculated the coronal hole deflection parameters, which are smaller for MCs.
The non-radial eruption and deflection of the CMEs has caused the FR eruptions originating from near DC to propagate away from DC and the observing spacecraft passed only by their flank and missed the central FR structures.
In addition, the MC-CMEs were also found to be wider than the EJ-CMEs.
The mean broadside and edge-on widths of the MC-CMEs 
were larger than those of the EJ-CMEs 
by  2 $^\circ$ and 7 $^\circ$, respectively. The obtained results suggest that the FR structures in the MC-CMEs are caught {\it in-situ} not only because they were deflected toward DC, but also because they were larger in size.

In conclusion, both the MC-CMEs and EJ-CMEs had similar solar sources and possible flux rope structures;
their different propagation directions and angular widths determine whether they are viewed 
as clouds or non-clouds by the observing spacecraft.
The results of this work support the conjecture that ``all CMEs have flux-rope structure".


\begin{acks}
This work uses data from the NASA/LWS Coordinated Data Analysis
Workshops on CME/flux-ropes in 2010 and 2011. We acknowledge the workshop support provided
by NASA/LWS, Predictive Sciences, Inc. (San Diego, CA), University of Alcala (Alcala
de Henares, Spain), and Ministerio de Ciencia e Innovacion (Reference number AYA2010-
12439-E), Spain. The authors acknowledge support of NASA grant LWSTRT08-0029.
\end{acks}

   


\begin{landscape}

\renewcommand{\thefootnote}{\alph{footnote}}

\begin{longtable}{lcccccccccccr}
\caption[List of the CDAW 2011 CMEs and the Best-fit Flux-rope Parameters.]
{List of the CDAW 2011 CMEs and the Best-fit Flux-rope Model Parameters.
}\label{T-FRfit}\\

\hline \multicolumn{4}{c}{ICME} &\multicolumn{4}{c}{CME} 
     & \multicolumn{5}{c}{Flux-rope Model Fit Output}\\ 
No. & Date & Time & \footnotemark[1]Type & Date & Time & $V_{sky}$ & Source &  $\omega_{e}$ &
$\omega_{b}$ 
     & $V_{cme}$ & Direction & \footnotemark[2]Tilt\\
    &   & (UT)  &   &  & (UT) & ($kms^{-1}$) &  & ($\circ$)&  ($\circ$)& ($kms^{-1}$) &  & ($\circ$) \\
\hline
\endhead

\hline\multicolumn{13}{|r|}{{Continued on next page}} \\ \hline
\endfoot

\hline \hline
\endlastfoot

01 & 1997/01/10 & 00:52 & MC & 1997/01/06 & 15:10 &    136 &  S18E06 & 55 & 70 &     258 &    S18W01 &  -60.00 \\
02 & 1997/05/15 & 01:15 & MC & 1997/05/12 & 05:30 &    464 &  N21W08 & 53 & 70 &     670 &    N01W02 &  -80.00 \\
04 & 1998/05/03 & 17:00 & EJ & 1998/05/02 & 23:40 &    585 &  S18W05 & 79 & 90 &     826 &    S16E14 &   42.00 \\
05 & 1998/05/04 & 02:00 & EJ & 1998/05/02 & 14:06 &    938 &  S15W15 & 79 & 90 &    2097 &    N08W05 &   22.00 \\
07 & 1998/11/07 & 08:00 & EJ & 1998/11/04 & 07:54 &    523 &  N17W01 & 71 & 95 &     706 &    N25W01 &   92.00 \\
08 & 1998/11/13 & 01:40 & EJ & 1998/11/09 & 18:18 &    325 &  N15W05 & 53 & 70 &     712 &    N15W05 &   16.00 \\
09 & 1999/04/16 & 11:10 & MC & 1999/04/13 & 03:30 &    291 &  N16E00 & 79 & 90 &     560 &    S02W06 &  -10.00 \\
10 & 1999/06/26 & 19:25 & EJ & 1999/06/24 & 13:31 &    975 &  N29W13 & 71 & 90 &    1531 &    N25W15 &   45.00 \\
13 & 1999/09/22 & 12:00 & EJ & 1999/09/20 & 06:06 &    604 &  S20W05 & 31 & 70 &     868 &    S20W05 &  -35.00 \\
14 & 1999/10/21 & 02:13 & EJ & 1999/10/18 & 00:06 &    144 &  S30E15 & 48 & 90 &     217 &    S30E15 &   60.00 \\
15 & 2000/01/22 & 00:23 & EJ & 2000/01/18 & 17:54 &    739 &  S19E11 & 71 & 90 &    1179 &    S10E29 &  -25.00 \\
16 & 2000/02/20 & 21:00 & MC & 2000/02/17 & 21:30 &    728 &  S29E07 & 71 & 84 &     994 &    S12W02 &   70.00 \\
17 & 2000/07/10 & 06:00 & EJ & 2000/07/07 & 10:26 &    453 &  N04E00 & 48 & 90 &     739 &    S17W05 &   70.00 \\
18 & 2000/07/11 & 11:22 & EJ & 2000/07/09 & 23:50 &    483 &  N18W12 & 58 & 77 &    1152 &    N18W06 &   15.00 \\
19 & 2000/07/15 & 14:18 & MC & 2000/07/14 & 10:54 &   1674 &  N22W07 & 90 & 16 &    2281 &    N18W14 &   30.00 \\
20 & 2000/07/26 & 18:58 & EJ & 2000/07/23 & 05:30 &    631 &  S13W05 & 71 & 84 &    1119 &    S13E04 &  -15.00 \\
21 & 2000/07/28 & 06:39 & MC & 2000/07/25 & 03:30 &    528 &  N06W08 & 64 & 84 &     960 &    S15E04 &  -95.00 \\
23 & 2000/08/11 & 18:51 & MC & 2000/08/09 & 16:30 &    702 &  N20E12 & 53 & 77 &    1024 &    N17E05 &  -85.00 \\
24 & 2000/09/17 & 17:00 & MC & 2000/09/16 & 05:18 &   1215 &  N14W07 & 64 & 90 &    1574 &    N08W07 &   45.00 \\
25 & 2000/10/05 & 03:23 & EJ & 2000/10/02 & 03:50 &    525 &  S09E07 & 53 & 90 &    1104 &    S19E08 &  -65.00 \\
26 & 2000/10/12 & 22:36 & MC & 2000/10/10 & 23:50 &    798 &  N01W14 & 58 & 95 &    1287 &    N20W14 &  -55.00 \\
27 & 2000/11/06 & 09:20 & MC & 2000/11/03 & 18:26 &    291 &  N02W02 & 64 & 87 &     542 &    N02E05 &   25.00 \\
28 & 2000/11/26 & 05:30 & EJ & 2000/11/24 & 05:30 &   1289 &  N20W05 & 79 & 95 &    1745 &    N30W18 &  -55.00 \\
29 & 2001/03/03 & 11:30 & EJ & 2001/02/28 & 14:50 &    313 &  S17W05 & 71 & 90 &     522 &    S05W15 &  -65.00 \\
30 & 2001/03/22 & 14:00 & EJ & 2001/03/19 & 05:26 &    389 &  S20W00 & 71 & 90 &     691 &    N05W10 &   85.00 \\
31 & 2001/04/11 & 14:12 & EJ & 2001/04/09 & 15:54 &   1192 &  S21W04 & 79 & 84 &    1813 &    S12E01 &   75.00 \\
32 & 2001/04/11 & 16:19 & MC & 2001/04/10 & 05:30 &   2411 &  S23W09 & 71 & 95 &    3735 &    S23W05 &  -85.00 \\
33 & 2001/04/28 & 05:02 & MC & 2001/04/26 & 12:30 &   1006 &  N20W05 & 79 & 77 &    1093 &    N20W03 &   30.00 \\
34 & 2001/08/12 & 11:10 & EJ & 2001/08/09 & 10:30 &    479 &  N11W14 & 45 & 90 &     842 &    N02W18 &   80.00 \\
35 & 2001/10/11 & 16:50 & EJ & 2001/10/09 & 11:30 &    973 &  S28E08 & 58 & 84 &    1449 &    S28E01 &  -20.00 \\
36 & 2002/03/18 & 13:13 & MC & 2002/03/16 & 23:06 &    957 &  S08W03 & 79 & 00 &    1151 &    N15W01 &  -50.00 \\
37 & 2002/04/17 & 11:01 & MC & 2002/04/15 & 03:50 &    720 &  S15W01 & 79 & 84 &    1302 &    S01W05 &  -10.00 \\
38 & 2002/05/11 & 10:30 & EJ & 2002/05/08 & 13:50 &    614 &  S12W07 & 48 & 90 &    1231 &    S09W09 &   55.00 \\
39 & 2002/05/18 & 19:51 & MC & 2002/05/16 & 00:50 &    600 &  S23E15 & 48 & 84 &     900 &    S23E05 &  -70.00 \\
40 & 2002/05/20 & 03:40 & EJ & 2002/05/17 & 01:27 &    461 &  S20E14 & 31 & 66 &     743 &    S28E20 &  -60.00 \\
41 & 2002/05/30 & 02:15 & EJ & 2002/05/27 & 13:27 &   1106 &  N22E15 & 58 & 84 &    1362 &    N32E20 &   80.00 \\
42 & 2002/07/17 & 15:50 & EJ & 2002/07/15 & 21:30 &   1300 &  N19W01 & 90 & 95 &    2046 &    N29E15 &  -40.00 \\
43 & 2002/08/01 & 05:10 & MC & 2002/07/29 & 12:07 &    222 &  S10W10 & 64 & 84 &     448 &    S02W10 &  -70.00 \\
44 & 2003/08/17 & 13:40 & MC & 2003/08/14 & 20:06 &    378 &  S10E02 & 48 & 84 &     662 &    N12E10 &  -65.00 \\
45 & 2003/10/29 & 06:00 & MC & 2003/10/28 & 11:30 &   2459 &  S16E08 & 90 & 90 &    2916 &    S16E20 &   75.00 \\
46 & 2003/10/30 & 16:20 & MC & 2003/10/29 & 20:54 &   2029 &  S15W02 & 90 & 95 &    3474 &    S15E05 &   80.00 \\
47 & 2004/01/22 & 01:10 & EJ & 2004/01/20 & 00:06 &    965 &  S13W09 & 90 & 71 &    1441 &    S25W10 &   60.00 \\
48 & 2004/07/24 & 05:32 & MC & 2004/07/22 & 08:30 &    700 &  N04E10 & 71 & 84 &    1359 &    N06E05 &  -10.00 \\
49 & 2004/11/09 & 09:05 & MC & 2004/11/06 & 02:06 &   1111 &  N09E05 & 90 & 95 &    1319 &    N07W00 &   12.00 \\
50 & 2004/12/11 & 13:03 & EJ & 2004/12/08 & 20:26 &    611 &  N05W03 & 79 & 77 &     754 &    S05W06 &   45.00 \\
51 & 2005/01/16 & 09:27 & EJ & 2005/01/15 & 06:30 &   2049 &  N16E04 & 58 & 84 &    2503 &    N25W01 &  -80.00 \\
52 & 2005/02/17 & 21:59 & EJ & 2005/02/13 & 11:06 &    584 &  S11E09 & 60 & 84 &     587 &    S21E19 &   75.00 \\
53 & 2005/05/15 & 02:19 & MC & 2005/05/13 & 17:12 &   1689 &  N12E11 & 90 & 90 &    2384 &    N05E11 &   45.00 \\
54 & 2005/05/20 & 03:34 & MC & 2005/05/17 & 03:26 &    449 &  S15W00 & 64 & 84 &     596 &    N08E05 &   85.00 \\
56 & 2005/07/10 & 02:56 & EJ & 2005/07/07 & 17:06 &    683 &  N09E03 & 48 & 87 &    1040 &    N12E26 &   91.00 \\
57 & 2005/09/02 & 13:32 & EJ & 2005/08/31 & 11:30 &    825 &  N13W13 & 58 & 90 &    1161 &    N08W25 &   -5.00 \\
58 & 2005/09/15 & 08:25 & EJ & 2005/09/13 & 20:00 &   1866 &  S09E10 & 79 & 95 &    2171 &    S29E21 &  -52.00 \\
59 & 2006/08/19 & 10:51 & EJ & 2006/08/16 & 16:30 &    888 &  S16W08 & 71 & 90 &    1351 &    S28W01 &  -15.00 \\
\footnotetext{\bf {Notes:}}
\footnotetext{Columns 1 - 4: the event number, shock date, time and ICME type. Columns 5 - 8: CME date, first appearance time at LASCO C2, 
              sky-plane speed and source location. Columns 9 - 13: flux-rope model fit edge-on width, broadside width, radial speed, 
              propagation direction and tilt angle.}\\ 

\footnotetext[1] {MC = Magnetic cloud; EJ = Ejecta} 
\footnotetext[2] {With respect to East clockwise}

\end{longtable}

\renewcommand{\thefootnote}{\arabic{footnote}}
\end{landscape}

\end{article} 

\end{document}